\documentclass[pra,twocolumn,aps,superscriptaddress,showpacs]{revtex4-1}
\usepackage{hyperref}
\usepackage{graphicx}
\usepackage{amsmath}
\usepackage{amsfonts}
\usepackage{amssymb}
\usepackage{epsfig}
\usepackage{lineno}
\usepackage{subfigure}
\usepackage{mathtools}
\usepackage[utf8]{inputenc}
\usepackage{braket}
\usepackage[usenames,dvipsnames]{color}
\usepackage{setspace}
\usepackage{bm}
\usepackage{times}
\hypersetup{
      colorlinks=true,
      citecolor=blue,
      linkcolor=blue,
      urlcolor=blue}

\begin{document}
\title{Spin-orbit coupling driven superfluid states in optical lattices 
       at zero and finite temperatures}
\author{Kuldeep Suthar}
\affiliation{Institute of Atomic and Molecular Sciences, 
             Academia Sinica, Taipei 10617, Taiwan}
\author{Pardeep Kaur}
\affiliation{Indian Institute of Technology Ropar,
             Rupnagar - 140001, Punjab, 
	     India}
\author{Sandeep Gautam}
\affiliation{Indian Institute of Technology Ropar,
             Rupnagar - 140001, Punjab, 
	     India}
\author{Dilip Angom}
\affiliation{Physical Research Laboratory,
             Ahmedabad - 380009, Gujarat,
             India}
\affiliation{Department of Physics, Manipur University,
             Canchipur - 795003, Manipur, India}
\date{\today}


\begin{abstract}
We investigate the quantum phase transitions of a two-dimensional Bose-Hubbard 
model in the presence of a Rashba spin-orbit coupling with and without thermal 
fluctuations. The interplay of single-particle hopping, strength of spin-orbit 
coupling, and interspin interaction leads to superfluid phases with distinct 
properties. With interspin interactions weaker than intraspin interactions, 
the spin-orbit coupling induces two finite-momentum superfluid phases. One of 
them is a phase-twisted superfluid that exists at low hopping strengths and 
reduces the domain of insulating phases. At comparatively higher hopping 
strengths, there is a transition from the phase-twisted to a finite momenta 
stripe superfluid. With interspin interactions stronger than the intraspin 
interactions, the system exhibits phase-twisted to ferromagnetic phase 
transition. At finite temperatures, the thermal fluctuations destroy the 
phase-twisted superfluidity and lead to a wide region of normal-fluid states. 
These findings can be observed in recent quantum gas experiments with 
spin-orbit coupling in optical lattices.
\end{abstract}

\maketitle

\section{Introduction}
The spin-orbit interaction plays a key role in several areas of condensed 
matter physics and material science like topological insulators and 
superconductors~\cite{hasan_10,qi_11,manchon_15}, quantum Hall 
effects~\cite{ezawa_13}, spin liquids~\cite{balents_10}, Weyl 
semimetal~\cite{wan_11}, and spintronics based devices~\cite{zutic_04}. Recent 
advances in ultracold quantum gas experiments have allowed the implementation 
of spin-orbit coupling (SOC) and competing interactions in strongly-correlated 
many-body systems~\cite{wu_16,sun_18,zhang_19}. These experimental developments
afford the possibilities to study novel states of matter, phase transitions, 
and exotic spin models which are not accessible in conventional condensed 
matter systems~\cite{radic2012exotic,cole_12}. The ultracold atomic systems 
are ideal platforms for such studies due to the tunability of potentials and 
multiparticle interactions.

In condensed matter systems, the SOC is an intrinsic property and can not be 
tuned~\cite{qi_11}. In contrast, it is possible to vary the strength of 
synthetic SOC in ultracold atoms by tuning the Raman coupling between 
pseudospin states and thereby different phase transitions can be 
explored~\cite{lin_11,li_16,li_17}. These experiments consider an 
equal-strength mixture of Rashba and Dresselhaus SOC. However, the experimental 
schemes to realize pure Rashba SOC in ultracold quantum gases have also been 
proposed~\cite{galitski_13}. A spin orbit-coupled pseudospin-$1/2$ Bose gas 
undergoes two successive magnetic phase transitions as the strength of Raman 
coupling is increased. First transition is from a stripe to a magnetized 
plane-wave phase, and second is from the magnetized plane-wave to a 
non-magnetic zero-momentum superfluid state for the Raman coupling of the 
order of the recoil energy~\cite{ji_14}. Furthermore, a two-dimensional (2D) 
SOC exhibits inversion and $C_4$ symmetries, thus opening new avenues to study 
topological band structures and quantum effects. The interaction-driven quantum
phase transition and topological region of such a 2D system are explored in a 
recent experiment~\cite{sun_18}. These experimental advances have led to 
several theoretical studies. The magnetic ordering such as spin-spiral 
ordering~\cite{cai_12}, vortex and Skyrmion crystal~\cite{cole_12}, and 
ferromagnetic and antiferromagnetic phases~\cite{gong_15} have been examined. 
The effects of strength and symmetry of SOC on the 
ground-state~\cite{bolukbasi_14}, crystal momentum 
distributions~\cite{yamamoto_17}, and SOC-driven Mott insulator (MI) to 
superfluid (SF) phase transition~\cite{yan_17} have also been investigated. 
The strong Rashba SOC destroys insulating domain and generates finite-momentum 
and stripe ordered superfluids~\cite{grass_11,mandal_12,cole_12}. The 
SOC-driven twisted superfluid states of binary spin mixtures in hexagonal 
optical lattice are also observed in quantum gas experiment~\cite{soltan_12}. 
The introduction of optical lattice potential breaks the Galilean 
invariance~\cite{hamner_15} and enhances the contrast, life-time, and parameter
regime of the stripe superfluid state~\cite{bersano_19}. Inspite of several 
studies, the parameter regions of the superfluid states in spin-orbit coupled 
Bose-Hubbard model and the effects of thermal fluctuations on the transition 
between finite-momentum superfluids have not been investigated. At finite 
temperatures, the melting of stripe superfluid phase leads to a wide domain of 
stripe normal-fluid (NF) phase~\cite{hickey_14}. More recently, it has been 
shown that the SOC leads to the lowering of the critical temperature for the 
superfluid to NF phase transition, and reduces the coherence and spatial orders
of magnetic textures~\cite{dutta_19}. 

In the present work, we study the ground-state phase diagrams of two-component 
interacting bosons in the presence of Rashba SOC at zero and 
finite temperatures. We examine the quantum phase transitions and characterize 
the SOC-driven finite-momentum superfluid states in two different regimes based
on the interspin interactions. For interspin interactions weaker than the 
intraspin interactions, the two finite momenta superfluids are phase-twisted 
(PT) and stripe (ST) superfluids whereas for interspin interactions stronger 
than intraspin interactions, these are PT and $z$-polarized ferromagnetic (zFM) 
superfluid phase. At $T = 0$K, the critical hopping of MI to PT superfluid 
transition decreases with SOC, which is in agreement to the mean-field 
predictions. In contrast to the condensed matter systems, the phase diagrams of
the system can be explored by tuning the experimental parameters. Furthermore, 
we extend our study to the case of finite-temperature, and show the interplay 
of SOC and thermal fluctuations on the superfluid states. We observe the 
melting of PT superfluid state into insulator and NF phase at finite 
temperatures. 

The paper is structured as follows: we introduce the model Hamiltonian of the 
present study and provide a brief description of the mean-field Gutzwiller 
approach in section~\ref{theory_gw_so}. In section~\ref{phase_char}, we 
provide the characterization of the superfluid states of the model considered. 
In section~\ref{results}, we first discuss the zero temperature phase diagrams 
of the Bose-Hubbard model in the presence of synthetic SOC, and then we examine 
the effects of finite-temperature on the SOC-driven superfluid states. 
Finally we conclude our findings in section~\ref{conc}.


\section{Model and Method}
\label{theory_gw_so}
We consider a pseudospinor system of ultracold bosons loaded into a square 
optical lattice. The two different atomic hyperfine levels of same atomic 
species act as two pseudospin states. The system is well described by a 
two-component Bose-Hubbard model (BHM) in the presence of Rashba SOC on a 
2D optical lattice. The Hamiltonian of the system is~\cite{yan_17}
\begin{eqnarray}
 \hat{H} = &-& J \sum_{\langle i,j \rangle} \hat{{\Psi}}^{\dagger}_{i}
           \hat{\Psi}_{j} + \sum_{i,\alpha} \left(\epsilon_{i\alpha} 
	- \mu \right)\hat{n}_{i\alpha} \nonumber \\
     &+& \frac{1}{2} \sum_{i,\alpha} U_{\alpha}~
         \hat{n}_{i\alpha}(\hat{n}_{i\alpha} - 1) 
      + U_{\uparrow\downarrow} \sum_{i} \hat{n}_{i\uparrow}
        \hat{n}_{i\downarrow} \nonumber \\ 
     &+& i\lambda \sum_{\langle ij \rangle} \hat{{\Psi}}^{\dagger}_{i}
	\hat{\bf{z}}\cdot\left(\vec{\mathbf \sigma}
	\times\vec{{d}_{ij}}\right)\hat{\Psi}_{j} 
     + {\rm H.c.},
\label{ham}
\end{eqnarray}
where $i$ is a unique combination of lattice site indices in 2D, i.e. 
$i\equiv(p,q)$ with $p$ and $q$ as site indices along $x$ and $y$ directions, 
respectively, and $j\equiv(p',q')$ is a neighboring site of $i$th site. Here 
$\Psi_{i} = (\hat{b}_{i\uparrow},\hat{b}_{i\downarrow})^{T}$ is a two-component
bosonic annihilation operator at $i$th lattice site, 
$\alpha = {\uparrow,\downarrow}$ denotes the pseudospin components, $J$ is 
spin-independent hopping amplitude of atoms, and for the present study, we 
consider equal hopping amplitudes for both the components, $\epsilon_{i\alpha}$
is the energy offset of atoms with $\alpha$ spin due to envelope confining 
potential and is considered to be zero, $\mu$ is the chemical potential, 
$\hat{n}_{i\alpha} = \hat{b}^{\dagger}_{i\alpha} \hat{b}_{i\alpha}$ is the 
number operator, and $U_{\alpha} (U_{\uparrow\downarrow})$ is intra-(inter-) 
spin on-site interaction. For the present work, we choose the intraspin 
interactions to be same, $U_{\uparrow} = U_{\downarrow} = U$. 
We consider $U$ as the scaling parameter for the tunneling
amplitude, chemical potential, interspin interaction, and the energy of 
the system. The last term represents the SOC generated by Raman lasers which 
can be tuned in experiments using coherent destructive hopping 
methods~\cite{zhang_13} and represents the hopping between neighbouring sites 
with a spin flip. Here $\lambda$ is the Rashba SOC strength, 
$\vec{\bf{\sigma}} = (\sigma_x,\sigma_y,\sigma_z)$ is a vector of Pauli spin 
matrices, $\vec{d}_{ij}$ is a lattice unit vector between two neighbouring 
sites, and $\hat{\bf{z}}$ is a unit vector perpendicular to the lattice plane.

To study the ground state properties of the system in both strong and weak 
coupling limit, we use single-site Gutzwiller mean-field (SGMF)
theory~\cite{gutzwiller_63,rokshar_91,krauth_92,sheshadri_93, menotti_07,
iskin_11,bandyopadhyay_19,suthar_20a}. In this theory, 
the many-body ground state is the product of single-site states. The Gutzwiller
ansatz is
\begin{equation}
  \ket{\Psi} = \prod_{i}\ket{\psi}_{i} 
	     = \prod_{i} \left(\sum^{N_b}_{n_{\uparrow},n_{\downarrow}} 
		c^{i}_{n_{\uparrow},n_{\downarrow}} 
		\ket{n_{\uparrow},n_{\downarrow}}_{i}\right), 
\end{equation}
where $\ket{\psi}_{i}$ is the single-site ground state, $N_b$ is the number of
occupation basis or maximum number of bosons corresponding to each spin state 
at each lattice site. Here $\ket{n_{\uparrow},n_{\downarrow}}_{i}$ is the 
occupation or Fock state, which is the direct product of the occupation states 
of both spin-components and $c^{i}_{n_{\uparrow},n_{\downarrow}}$ is the 
corresponding Gutzwiller coefficient of the coupled Fock state. 
The SF order parameter and the average occupancy are defined as  
$\phi_{i\alpha} = \langle\Psi|\hat{b}_{i\alpha}|\Psi\rangle$ and 
$n_{i\alpha} = \langle\Psi|\hat{n}_{i\alpha}|\Psi\rangle$, respectively.
This mean-field approach has been employed in the studies of bosons in optical 
lattices with synthetic magnetic field and 
SOC~\cite{kuno_17,bai_18,pal_19,suthar_20,yan_17}. To study the effects of 
thermal fluctuations at finite temperatures, we use the finite-temperature 
Gutzwiller theory, and a brief description is presented in the appendix.

We analyze the system in weakly-interacting limit, where $U_{\alpha}\ll J$. For
this regime, the Hamiltonian, Eq.~(\ref{ham}), in the momentum space can be 
written as 
\begin{equation}
 \hat{H}_{\rm kin} = \sum_{\mathbf k} \left(\begin{array}{cc}
 \hat{b}^{\dagger}_{k \uparrow} & 
 \hat{b}^{\dagger}_{k \downarrow} \end{array}\right) \mathcal{H}_{\mathbf k}
 \left(\begin{array}{c} \hat{b}_{k \uparrow}  \\ \hat{b}_{k \downarrow}
 \end{array}\right)
\end{equation}
with
\begin{equation}
  \mathcal{H}_{\mathbf k} = \left(\begin{array}{cc}
  -2 J (\cos k_{x} + \cos k_{y}) &
   2 i \lambda(\sin k_{x} - i \sin k_{y}) \\
  -2 i \lambda(\sin k_{x} + i \sin k_{y}) &
  -2 J (\cos k_{x} + \cos k_{y}) \end{array}\right).
 \nonumber \\
\end{equation}

\begin{figure}
  \includegraphics[width=\linewidth]{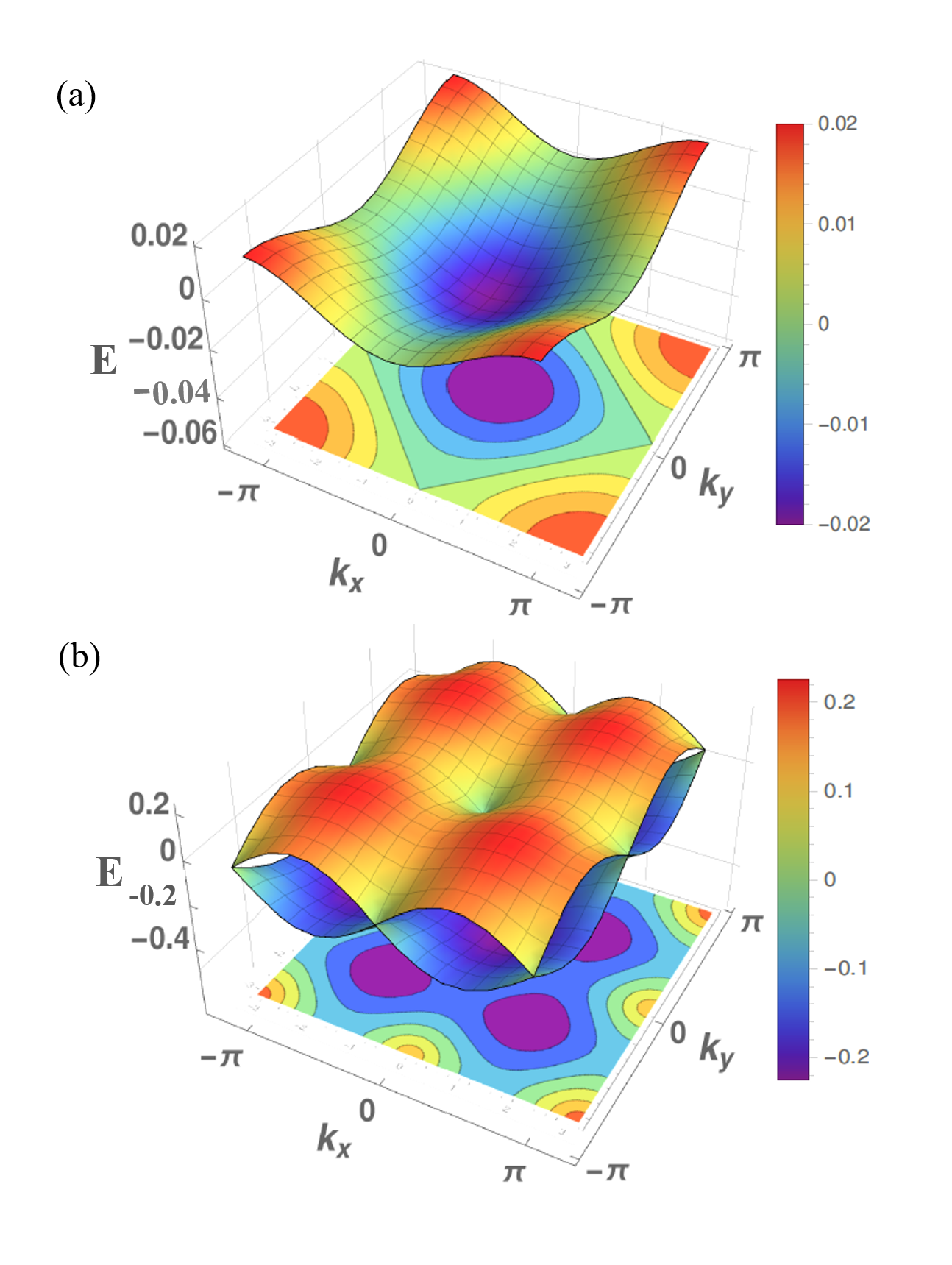} 
  \caption{Noninteracting band structure of two-dimensional square optical 
	   lattice for two regimes (a) $\lambda/J = 0$ representing the 
           single minima at $\mathbf{k} = 0$ in the absence of SOC. The 
           other case is shown for finite SOC, where the competition between 
           $\lambda$ and $J$ determines the band structure. 
           (b) $\lambda/J = 8$, the SOC breaks the rotational symmetry, and 
           the minima occur at four finite wave-vectors in the lower branch. 
           This is evident from the projection of lower energy branch onto the 
           $k_x$-$k_y$ plane. As $\lambda/J$ decreases, the minima of the lower 
           branch tends to approach ${\bf k}=0$.} 
 \label{band}
\end{figure}

The diagonalization of the above Hamiltonian $\mathcal{H}_{\mathbf k}$ yields 
two energy branches 
\begin{equation}
  E_{\mathbf{k}\pm} = - 2J\left(\cos k_{x} + \cos k_{y} \right)
               \pm 2 \lambda \sqrt{\sin^2 k_{x} 
               + \sin^2 k_{y}},
\end{equation}
where $\mathbf{k}=(k_x,k_y)$, and the first term is spin-independent dispersion 
relation in a 2D square lattice. The energy spectrum $E_{\mathbf{k}\pm}$ 
remains invariant under the parity transformation 
$(k_x \rightarrow -k_x, k_y \rightarrow -k_y)$ and permutation of $k_x$ and 
$k_y$, $(k_x \rightarrow k_y, k_y \rightarrow k_x)$. The non-interacting lowest
band structure is shown for two different regimes in Fig.~\ref{band}. In the 
absence of SOC $\lambda=0$, the lower branch of the band has one minimum at 
$\mathbf{k}=(0,0)$. The SOC term modifies the band structure, where the lower 
branch has four degenerate minima. The presence of SOC breaks the rotational 
symmetry in k-space and shifts the minima of the lower branch along the 
diagonals of the first Brillouin zone. 

The four degenerate minima in the lower branch are $\mathbf{q}_1 = (k_0,k_0), 
\mathbf{q}_2 = (-k_0,k_0), \mathbf{q}_3 = (-k_0,-k_0), 
\mathbf{q}_4 = (k_0,-k_0)$ where $k_0 = \arctan(\lambda/\sqrt{2}J)$. Hence, 
the location of the minima is determined by the strength of SOC. 


\section{Quantum phases and order parameters}
\label{phase_char}
 The ground states of ultracold bosons with SOC exhibit insulating and various 
SF phases. The nature of the SF phase depends on the competition between the 
single-particle hopping and SOC induced spin-dependent hopping. At lower $J$, 
the incompressible insulating phases are identified by the sum and difference 
of the expectations of the number operators, 
$\langle\hat{n}_{\pm}\rangle\equiv\langle\hat{n}_{\uparrow}\rangle
\pm\langle\hat{n}_{\downarrow}\rangle$. For the MI phase 
$\langle\hat{n}_{\pm}\rangle$ is an integer while the SF phases are 
characterized by real $\langle\hat{n}_{\pm}\rangle$ and finite value of the 
compressibility $\kappa=\partial\langle \hat{n}\rangle/\partial\mu$. In the 
absence of SOC, the amplitude and phase of the order parameter 
$\phi_{\uparrow}(\phi_{\downarrow})$ are homogeneous. 

The striking features appear when the spin-dependent hopping due to the SOC is 
finite. This is a complex hopping, it flips the atomic spin while hopping and 
causes variations in the phase of SF states. To classify the various SF states 
which feature distinct phase distributions of the order parameter, we examine 
the spin-dependent momentum distributions at the wave vector $\mathbf{k}$
\begin{equation}
  \langle \rho_{\uparrow,\downarrow}(\mathbf{k}) \rangle = {L}^{-2} 
	                 \sum_{i,j} \langle \hat{b}^{\dagger}_{i\uparrow} 
			 \hat{b}_{j\downarrow}\rangle 
			 e^{i\mathbf{k}\cdot(\mathbf{r}_{i} - \mathbf{r}_{j})},
\end{equation}
where $L$ is the system size and  $\mathbf{r}_{i}$ ($\mathbf{r}_{j}$) is the 
location of $i$th ($j$th) lattice site. When the interspin interaction is
weaker than intraspin, the SF state can be of three types. These are (i) 
homogeneous superfluid: this has uniform amplitude and phase of the order 
parameter. For this state, the condensation occurs at zero momentum and is 
also referred as zero-momentum SF (ZM-SF) state. (ii) Phase-twisted (PT) 
superfluid: for this state the amplitude of 
$\langle \hat{b}_{i\alpha} \rangle$ is uniform but the phase varies
diagonally across the lattice (iii) Stripe (ST) superfluid: this state has 
stripe-like variation in the phase of $\langle \hat{b}_{i\alpha} \rangle$ across the 
lattice. Thus, we distinguish superfluid states based on their phase variation 
and momentum distributions. It is worth mentioning, similar SF states have 
been previously discussed in the continuum where the phases were characterized 
using the properties of collective excitations~\cite{chen_17}.

The interplay of spin-dependent hopping (SOC) and single-particle hopping leads 
to the exotic SF states. We examine the spin-dependent momentum 
distributions $\langle \rho_{\uparrow,\downarrow}(\mathbf{k}) \rangle$ at 
$\mathbf{k}=0$, $\langle \rho_{\uparrow,\downarrow}(\pm k_0, 0) \rangle$, 
$\langle \rho_{\uparrow,\downarrow}(0, \pm k_0) \rangle$, and 
$\langle \rho_{\uparrow,\downarrow}(\mathbf{q}_i) \rangle$. Here, 
$\mathbf{q}_i$ and $k_0$ depend on the ratio of the hopping to the SOC strength
as discussed in Sec.~\ref{theory_gw_so}. For the PT superfluid, the momentum 
distribution at $\langle \rho_{\uparrow,\downarrow}(\mathbf{q}_i) \rangle$ is 
finite either at all the $\mathbf{q}_i$'s or only at one of the 
$\mathbf{q}_i$'s. This is due to the variation in phase distributions along 
the diagonal. Hence, this state shows a peak along the diagonal of the 
Brillouin zone in the $\mathbf{k}$-space. On the other hand, for the ST 
superfluid states, depending on the phase variation being horizontal or 
vertical, the state exhibits peak at 
$\langle \rho_{\uparrow,\downarrow}(\pm k_0, 0) \rangle$ or 
$\langle \rho_{\uparrow,\downarrow}(0, \pm k_0) \rangle$, respectively.
We define $\Phi = \langle \rho_{\uparrow,\downarrow}(k_0, 0) \rangle
+ \langle \rho_{\uparrow,\downarrow}(-k_0, 0) \rangle 
+ \langle \rho_{\uparrow,\downarrow}(0, k_0) \rangle
+ \langle \rho_{\uparrow,\downarrow}(0, -k_0) \rangle$, which serve as an 
order parameter to identify the PT-ST phase transition. Here, $\Phi$ is zero 
for the PT superfluid while finite for the ST state. As both PT and ST are 
SOC-driven finite-momentum superfluid states, therefore 
$\langle \rho_{\uparrow,\downarrow}(0,0) \rangle = 0$. In the next section, 
we have characterized the various phase transitions and the finite-momentum 
superfluids based on the aforementioned classification. Furthermore, when 
the interspin interaction is strong $U_{\uparrow\downarrow}>1$, we also report 
a ferromagnetic phase where the spins orient along $\pm z$ axis. This is 
referred to as $z$-polarized ferromagnetic (zFM) superfluid state where 
$\langle \hat{b}_{i\uparrow}\rangle$ ($\langle \hat{b}_{i\downarrow}\rangle$) 
remains finite and homogeneous but $\langle \hat{b}_{i\downarrow}\rangle$ 
($\langle \hat{b}_{i\uparrow}\rangle$) vanishes throughout the lattice. This phase 
can be easily distinguished from other superfluid states with finite 
$\phi_{\uparrow}$ or $\phi_{\downarrow}$. 

At finite temperatures, the superfluid states are characterized in a similar 
way as in the case of zero temperature, although the observables are defined 
with the thermal averages. The definitions of the thermal-averaged SF order 
parameter and occupancy are provided in the appendix. The normal-fluid state 
at finite temperatures is identified in the incompressible phases based on 
the compressibility $\kappa$. In the present work, we consider 
$|n-n_{\rm th}|\geqslant 10^{-3}$ as the criterion to identify MI to NF 
crossover. Here, $n$ and $n_{\rm th}$ are the lattice occupancies  at zero and 
finite temperatures, respectively. Such a criterion has been previously used 
to distinguish the NF phase of Bose-Hubbard model at finite 
temperatures~\cite{buonsante_04}.


\section{Results and discussion}
\label{results}
 We study the mean-field ground state phase diagram of the ultracold bosons 
and investigate the different SF phases emerging from  the competition between 
the SOC and single-particle hopping. In particular, we examine the system for 
weak ($U_{\uparrow\downarrow}/U < 1$) and strong 
($U_{\uparrow\downarrow}/U > 1$) interspin interactions. We, then, employ 
finite-temperature Gutzwiller theory to probe the effects of thermal 
fluctuations on the SF phases of the bosons. To generate the phase diagrams, 
we consider a system size of $8 \times 8$, and Fock state dimension at each 
lattice site is $N_b=6$. The latter is sufficient to represent the quantum 
phases of the system upto 
$\mu=3U$~\cite{gutzwiller_63,rokshar_91,krauth_92,sheshadri_93}. 
It is important to note that the initial states play a key role 
in determining the ground states. We have performed the numerical simulations 
with different initial states and found that a random SF order parameter as 
the initial state gives the global minima. The uniform $\phi$'s is not a good 
choice of the initial state because for some values of the parameters the 
converged solution corresponds to a local minima. This is due to the fact 
that the uniform $\phi$ do not contribute to the SOC energy as this depends on 
the relative phase between the $\phi$'s of both the pseudospinor components. 
To obtain the mean-field phase diagrams, we start with a complex random 
distribution of Gutzwiller coefficients across the lattice, and then the 
corresponding SF order parameter is computed. Hence, our initial state has a 
random SF order parameter with random amplitude and phase. Our algorithm is 
based on the self-consistent approach. We diagonalize the single-site 
Hamiltonians and compute the updated $\phi_{i\alpha}$'s at each iteration. 
This process is repeated until the energy and superfluid order parameter 
converge upto a tolerance of $10^{-12}$. Moreover, we repeat the procedure 
with $50$ random configurations of the initial state to ensure that the 
ground state has been obtained. We have checked explicitly that the 
larger number of random configurations do not modify the ground states.

\begin{figure}[ht]
  \includegraphics[width=\linewidth]{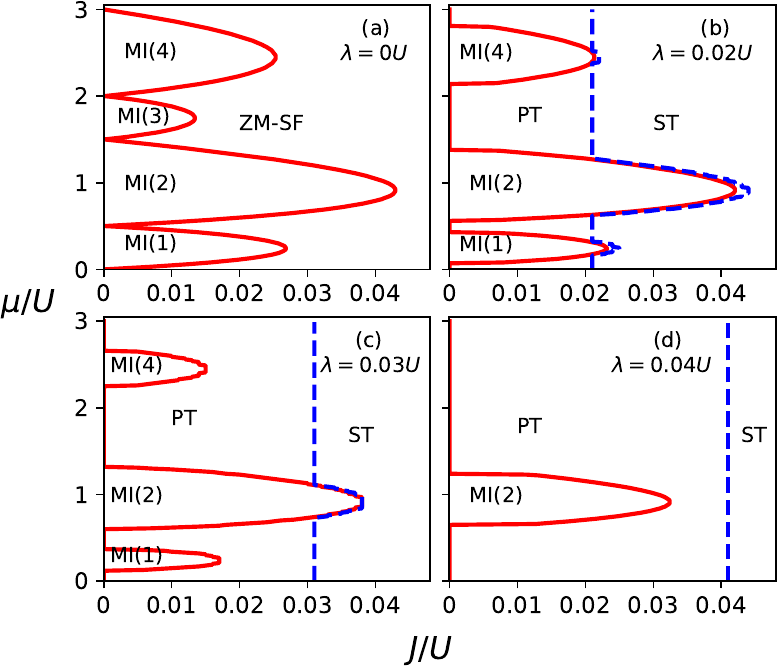}
  \caption{The zero-temperature ground state phase diagram in the presence of 
           Rashba SOC for various SOC strengths. The Mott insulator regime
	   is represented by MI$(n)$, where $n=n_{\uparrow}+n_{\downarrow}$ 
	   is the total filling or occupancy of the lobe. The interspin 
	   interaction $U_{\uparrow\downarrow} = 0.5U$, and the values of 
	   $\lambda$ in units of $U$ are shown in the upper right corner of 
	   the plot. At $\lambda=0$, the system exhibits MI-SF transition, 
           where the SF phase Bose condenses at zero momentum, and hence 
           referred to as the zero momentum (ZM) superfluid state. The 
           finite $\lambda$ results in finite-momentum superfluid phases. 
           Here, as $J$ is varied the system undergo PT-ST superfluid phase 
           transition, shown by blue dashed lines. The phase diagrams are 
           obtained using random complex initial states with $50$ random 
           configurations. The system size $L = 8\times 8$ and periodic boundary 
           conditions are considered.}
 \label{bhm_soc}
\end{figure}


\subsection{$U_{\uparrow\downarrow}=0.5$}
We first examine the quantum phases of 2D BHM in the presence of SOC 
interaction at zero temperature. The plots in the Fig.~\ref{bhm_soc} show the 
ground state phase diagrams at different values of the SOC strengths 
($\lambda$'s) with $U_{\uparrow\downarrow} = 0.5U$. 


\subsubsection{No spin-orbit coupling ($\lambda=0$)}
 In the absence of SOC, the system supports two quantum phases, the 
incompressible MI phase and the compressible ZM-SF phase. The MI phase occurs 
in lobes of different integer commensurate densities. It is to be noted that 
the MI lobes with odd integer occupancies are smaller than those with even 
occupancies~\cite{bai_20}. As $U_{\uparrow\downarrow}$ increases, the size of 
the odd-integered Mott lobes grow whereas even-integered lobes remain
same in size but shift to higher $\mu/U$ until $U_{\uparrow\downarrow} = U$. In 
the absence of SOC, the phase diagram shown in Fig.~\ref{bhm_soc}(a) agrees 
well with previous studies on the two-component BHM
~\cite{yamamoto_13,yamamoto_13a,kato_14,bai_20}. 


\subsubsection{Finite spin-orbit coupling ($\lambda\neq0$)}
  The ground state phase diagram for finite SOC are shown in 
Fig.~\ref{bhm_soc}(b)-(d). Consider the phase diagram at $\lambda = 0.02U$, a 
prominent feature is the shrinking of the MI lobes. At higher $\mu$, the 
MI(3) lobe vanishes and is replaced by the SOC-induced SF phase. Thus, even 
in the atomic limit $J/U=0$, for certain ranges of $\mu$, the system is in the 
SF phase due to the SOC. This is evident from the phase diagram in 
Fig.~\ref{bhm_soc}(b), where the SF phase is present at $J/U=0$ for 
$\mu/U\leqslant0.07, 0.43<\mu/U<0.57$, $1.38<\mu/U<2.14$ and 
$2.8<\mu/U\leqslant 3.0$. In the absence of single-particle 
hopping i.e. $J=0$, the superfluidity is due to the transport of atoms in the 
presence of spin-dependent hopping (SOC). As we increase $\lambda$, the MI 
lobes shrink further, and the SF phase is enhanced. For $\lambda=0.04 U$, only 
MI(2) lobe survives, and the system is in the SOC-generated SF phases in the 
remaining parameter domain. This is due to the relatively larger 
region covered by MI(2) lobe even at $\lambda=0$. Our computations for larger 
$\lambda$ show that the MI(2) lobe also vanishes at $\lambda\approx 0.06 U$. 
Hence, the MI lobe with larger insulating domain and higher $J_c$ will require 
larger SOC strengths to result in superfluid states occupying the whole domain 
of $J/U-\mu/U$ plane. The vanishing of insulating lobes with the formation of 
SOC-induced SF states is in agreement with the previous 
studies~\cite{mandal_12,yan_17}. Using site-decoupling approximation and 
second-order perturbation theory, the critical hopping of MI-SF transition in 
the presence of SOC is 
\begin{equation}
\left(\frac{zJ_c}{U}\right) = \frac{1}{2} 
	                      \left[\left(\frac{zJ_{0}}{U}\right) 
			    + \sqrt{\left(\frac{zJ_{0}}{U}\right)^{2} 
			    - 8\left(\frac{\lambda}{U}\right)^{2}} \right],
\label{jc}
\end{equation}
where $J_0$ is the critical hopping of MI-SF transition in the absence of 
SOC ($\lambda=0$). The value of $J_0$ depends on the occupation number of the 
species~\cite{guang_03}. Here $z=2d$ is the coordination number of the 
$d$-dimensional optical lattice. The value of $J_c$ decreases with $\lambda$, 
which confirms our numerical results in the phase diagrams shown in 
Fig.~\ref{bhm_soc}. As an illustration, for MI(2)-SF phase transition, the 
above expression yields the value of $J_c$ as $0.0418$ and $0.0402$ for 
$\lambda/U=0.02$ and $0.03$, respectively, which are in close agreement to the 
numerical values in phase diagrams in Figs.~\ref{bhm_soc}(b)-(c). 
Using Eq.~\ref{jc}, the critical SOC strengths where MI(2) and MI(4) lobes 
destroy are $0.061$ and $0.035$, respectively and these are consistent to our 
numerical results.

At $\lambda=0$, the only superfluid phase of the system is ZM 
superfluid, whereas at non-zero $\lambda$ value, it is replaced by 
finite-momentum superfluids. The nature of superfluid phases near the MI-SF 
transition can be understood by analyzing the mean-field 
energies~\cite{mandal_12}. The hopping energy depends on the relative phase 
between the same spin state while the SOC energy depends on the relative phase 
between different components. For $\lambda=0$ case, the minimization of hopping
energy leads to zero phase difference between the states which corresponds to 
ZM superfluid [Fig.~\ref{bhm_soc}(a)]. For finite $\lambda$, the energies 
depend on $\lambda/J$, relative phases, and the ratio of the amplitude of 
$\phi_{\downarrow}$ and $\phi_{\uparrow}$. For fixed $\lambda/J$ and assuming 
uniform amplitude of order parameters, the minimization of energies with 
respect to relative phases correspond to finite but uniform relative 
phases~\cite{mandal_12}, which is identified as PT superfluid state.

\begin{figure}[ht]
  \includegraphics[width=\linewidth]{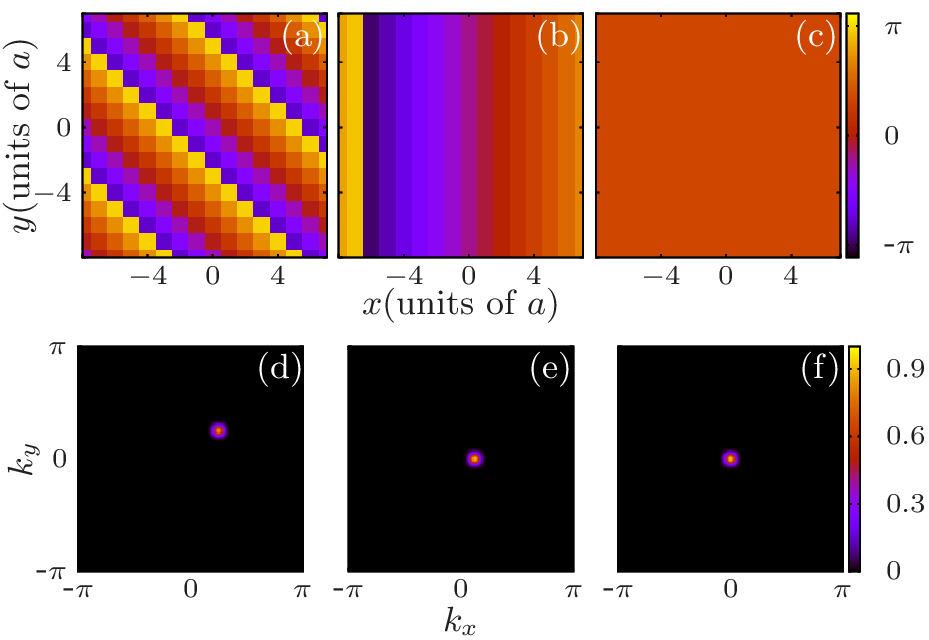}
  \caption{The lattice-site distributions of the phase variation and
           spin-dependent momentum distributions of various superfluid 
           states. The upper panel represents the phase 
	   distributions for (a) PT (b) ST (c) ZM superfluids. And, the 
	   momentum distributions are shown in the lower panel (d,e,f). The 
	   finite-momentum superfluids are obtained using the Gutzwiller 
	   mean-field approach for $\lambda = 0.02U$ and $\mu=1.5U$. The 
	   hopping amplitude in terms of $U$ corresponding to PT and ST are 
           $0.015$ and $0.04$, respectively. The ZM-superfluid (c) is plotted 
	   for $\lambda=0$, $\mu=1.5$, and $J=0.04U$. The spatial variation of 
	   phase and momentum distributions are shown for one of the 
	   components, as the other component also has the similar 
	   distributions. The peak in the spin-dependent momentum distributions 
           appears at $\mathbf{k}\neq 0$ for PT (d) and ST (e) states, whereas
           for ZM-superfluid it appears at $\mathbf{k}=0$ (f). Here, $a$ is the 
           lattice constant.}
 \label{soc_state}
\end{figure}

The characteristic properties of the finite-momentum and ZM-superfluid states 
are shown in Fig.~\ref{soc_state}. The phase variations and the momentum 
distributions of the finite-momentum superfluids are shown for fixed 
$\lambda=0.02U$, $\mu=1.5U$, and two different $J$ values corresponding to PT 
and ST superfluids. For the PT superfluid state, the random initial state 
yields uniform amplitude and twisted diagonal site-variation in the phase as 
evident from Fig.~\ref{soc_state}(a). The phase variation is shown for one of 
the components, although it is to be noted that the other component also 
follows similar distributions. However, the relative phase of the $\phi$'s 
between the components is finite, i.e. 
$\theta_{i\uparrow} \neq \theta_{i\downarrow}$~\cite{mandal_12}. In the 
presence of the interactions, the four-fold symmetry of lower branch of lowest 
energy band is spontaneously broken. For the PT superfluid phase, the system 
chooses to be in one of the minima and therefore, we observe a single peak at 
$\mathbf{k}\neq0$ in the momentum distribution. In particular, the peak in the 
$\mathbf{k}$-space appears at the diagonal of the Brillouin zone as represented
in Fig.~\ref{soc_state}(d). 

As $J$ is increased, PT phase undergoes a transition to the ST phase. In the ST
superfluid phase obtained with SGMF approach, the amplitude of $\phi_{i\alpha}$
remains spatially uniform and phase distributions exhibit stripe-like variation
[Fig.~\ref{soc_state}(b)]. The momentum peak is located at $\mathbf{k}\neq 0$, 
and in particular it lies on $x-$ or $y-$ axis depending on the variation in 
phase [Fig.~\ref{soc_state}(e)].

\begin{figure}[ht]
  \includegraphics[width=0.9\linewidth]{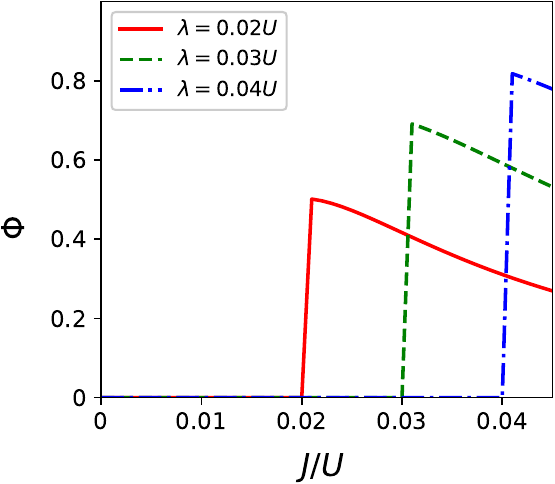}
  \caption{The evolution of the order parameter $\Phi$, characterizing the 
	   finite-momentum PT (phase-twisted) and ST (stripe) superfluid 
	   states, as a function of the hopping strength $J/U$. The chemical 
	   potential $\mu/U=1.8$ and interspin interaction 
           $U_{\uparrow\downarrow}=0.5 U$. The $\Phi$ is defined in 
	   Sec.~\ref{phase_char}. The variation in $\Phi$ from zero to finite 
	   shows PT(MI)-ST phase transition.}
 \label{op_lam}
\end{figure}

To examine the quantum phase transition between phase-twisted and ST superfluid
state, we analyze the properties of 
$\langle\rho_{\uparrow\downarrow}({\bf k})\rangle$. Since both states are 
finite-momentum superfluids, their location of momentum peaks in 
$\mathbf{k}$-space can serve as an order parameter to identify them. As 
mentioned earlier in Sec.~\ref{phase_char}, we in particular analyze the 
evolution of the order parameter $\Phi$, which is the sum of 
$\langle\rho_{\uparrow\downarrow}({\bf k})\rangle$ at $\mathbf{k}=(\pm k_0,0)$ 
and $(0,\pm k_0)$, as a function of $J$. For each $\mu$ value, we have spanned 
along the $J/U$-axis, and whenever $\Phi$ takes a non-zero value, critical 
hopping strength for PT-ST transition is determined. The error involved in 
analyzing the phase transition is $10^{-3}$ which is the step size $(\Delta J)$
used to span $J/U$ in the numerical computations. Hence, we find that the PT-ST
phase transition is sharp. As a representative case, the evolution of $\Phi$ 
at $\mu/U=1.8$ for three different $\lambda$ values are shown in 
Fig.~\ref{op_lam}. At lower hopping strengths, the ground state is either MI 
phase or the finite-momentum PT superfluid, and hence $\Phi$ remains zero. 
This is due to the fact that PT state corresponds to the condensation in 
${\bf k} = {\bf q_i}$ along diagonals of the first Brillouin zone. As $J/U$ 
increases, a striped-ordering of the phase develops with finite $\Phi$, this 
characterize PT-ST superfluid phase transition of the spin-orbit coupled 
bosons. The critical hopping strength of the PT-ST transition increases as the 
value of SOC strength increases. As shown in Fig.~\ref{op_lam}, the $J_c$ of 
PT-ST transition is $0.02$, $0.03$, and $0.04$ for $\lambda=0.02,0.03$ and 
$0.04$, respectively. The behaviour of $\Phi$ and the corresponding transitions
for $U_{\uparrow\downarrow} = 0.5$ [Fig.~\ref{bhm_soc}] suggest that the PT to 
ST superfluid phase transition occurs when $\lambda/J \approx 1$. And, the PT 
phase is expected for $\lambda/J \lessapprox 1$ whereas the ST phase appears 
for $\lambda/J \gtrapprox 1$. Our computations using random configurations of 
complex $\phi$'s suggest an increase in $J_c$ as the system size increases. 
Since the real cold-atom experiments are with the trapped finite-size systems, 
therefore the transition between finite-momentum superfluids can be observed 
near the trap-center~\cite{yan_17}.

\begin{figure}[ht]
  \includegraphics[width=\linewidth]{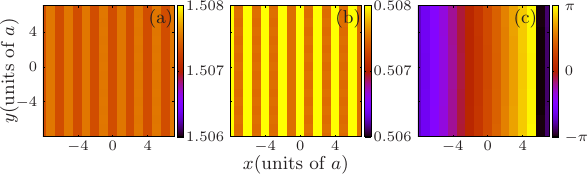}
  \caption{The lattice-site distributions of the $\rm ST_{den}$ phase 
           obtained using cluster Gutzwiller approach. (a) the occupancy
           (b) amplitude of the order parameter, and (c) phase of a 
	   $\rm ST_{den}$ state are shown. The parameters considered are 
	   $\lambda=0.02 U$, $\mu=1.8 U$, and $J=0.05 U$. These distributions 
	   are shown for interspin interaction $U_{\uparrow\downarrow}=0.5 U$.}
\label{st_cl}
\end{figure}

The SGMF approach failed to capture the density oscillations that should 
ideally be there in a stripe phase, and the reason is that SGMF does not 
include the inter-site atomic correlations. In order to overcome this 
limitation of SGMF, and obtain the nonuniform magnetic ordering and the 
resulting inhomogeneous superfluidity, one has to use the diagonalization of 
cluster of lattice sites as suggested in Ref.~\cite{cole_12}. Considering 
this, we probe the parameter space of the stripe superfluid state obtained 
from SGMF theory with the cluster Gutzwiller approach (CGA). The latter 
improves the inter-site correlations and incorporates the effects of the 
quantum fluctuations. In this approach, the lattice sites are partitioned 
into a finite number of clusters, where the model terms within the lattice 
sites of a cluster are treated exactly. The detailed description of the 
approach is given in our previous 
works~\cite{bai_18,pal_19,suthar_20,bai_20,suthar_20a}. To examine the 
parameter domain corresponding to the stripe superfluid, we use $2\times 2$ 
cluster and $N_{b} = 3$. The size of the clusters is sufficient to probe the 
effects of the atomic correlations on the magnetic ordering of the SOC-driven 
superfluids. Like in the case of SGMF, sometimes the solution obtained from 
CGA is a metastable state corresponding to a local minima. To avoid this we 
consider several random configurations of the SF order parameters as the 
initial states with CGA and choose the global minimum-energy state as the 
ground state phase. The lattice-site distributions of the occupancy, and the 
amplitude and phase of the SF order parameter are shown in Fig.~\ref{st_cl}. 
The profiles are shown for one of the component, $|\uparrow\rangle$, however 
the other component also follows the similar distributions. We observe the 
stripe variation in the number occupancy $\langle \hat{n}_{i\alpha} \rangle$ 
and $|\phi_{i\alpha}|$, and hence term it as ${\rm ST}_{\rm den}$ phase. The 
amplitude of the variations remain smaller, which we expect, can be enhanced 
by considering larger cluster of sites. 

We further investigate the parameter domain of the stripe superfluid 
(using CGA) and find that the ${\rm ST}_{\rm den}$ phase persists for larger 
hopping strengths. It continues to the domain where one would get ZM-SF 
transition using SGMF. As an example, for the parameters $L=8\times 8$, 
$\lambda=0.02U$, $\mu=1.8U$, and $J=0.1U$, the SGMF predicts ZM-superfluidity 
whereas the CGA gives ${\rm ST}_{\rm den}$ phase for these parameters. Hence, 
the latter extends the parameter space of ${\rm ST}_{\rm den}$ phase by taking 
into account the quantum correlations. This suggests the applicability of SGMF 
to describe the quantum phase transitions usually for 
$J/U \rightarrow 0$~\cite{cole_12,mandal_12,yan_17}. Therefore, for the 
present work, we have investigated the phase transitions in the range from 
$J=0$ to $J\approx 0.08 U$. The stability of the ${\rm ST}_{\rm den}$ 
superfluid in a wider parameter regime is consistent with the observation of 
this state in the presence of weak lattice potential in a recent 
experiment~\cite{bersano_19}. In the present work, the CGA has been used to 
ascertain the nature of ST and ZM phases obtained from SGMF. However, the 
detailed analysis of the phase diagrams with CGA can be addressed in a future 
work.

\subsection{$U_{\uparrow\downarrow}=1.5$}
\begin{figure}[ht]
  \includegraphics[width=\linewidth]{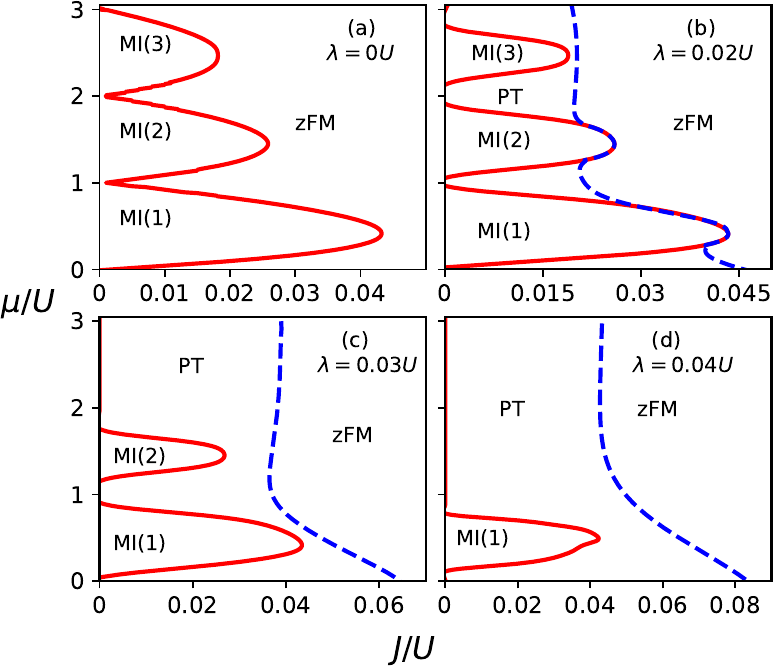}
  \caption{The zero-temperature ground state phase diagram in the presence of
           Rashba SOC for various SOC strengths. The filling or occupancy of 
	   the Mott lobe is represented by $n$ in MI$(n)$. The interspin 
	   interaction $U_{\uparrow\downarrow} = 1.5U$, and the values of 
	   $\lambda$ in units of $U$ are shown in the upper right corner of 
	   the plot. The superfluidity near the MI lobes is twisted in 
	   character while at higher $J$ the system is in zFM state.}
\label{pd_1p5}
\end{figure}
Here, we discuss the quantum phases generated due to the effects of SOC when 
the interspin interaction is stronger than the intraspin ones 
$(U_{\uparrow\downarrow}/U > 1)$. In this parameter regime, we first review 
the phase diagram of two-component interacting bosonic system in the absence 
of SOC. The phase diagram for $U_{\uparrow\downarrow} = 1.5 U$ at $\lambda=0$ 
is shown in Fig.~\ref{pd_1p5}(a). Above the phase separation criterion, at 
$J=0$, the width of all MI lobes is $\Delta\mu/U = 1$. Moreover, the critical 
hopping of the MI(1)-SF transition in Fig.~\ref{pd_1p5}(a) becomes identical 
to the MI(2)-SF for $U_{\uparrow\downarrow}< U$ case as shown in 
Fig.~\ref{bhm_soc}(a). The details of the quantum phase transitions as a 
function of $U_{\uparrow\downarrow}$ for two-component interacting 
scalar-bosonic system is reported in our previous study~\cite{bai_20}. In the 
phase-separated superfluid, the condensation occurs in one of the component 
only, and it resembles the zFM phase~\cite{cole_12,dutta_19}. We further 
examine the SOC-driven superfluid phases and their parameter space as 
$\lambda$ varied. The phase diagram for three representative cases are shown in 
Fig.~\ref{pd_1p5}(b)-(d). At $\lambda=0.02 U$, for lower hopping strengths, 
the phase-modulated PT superfluid emerges between the insulating lobes as 
shown in Fig.~\ref{pd_1p5}(b). For the phase-separated regime, 
the uniform occupancy of PT state is observed for 
$n_{i}=n_{i\uparrow}+n_{i\downarrow}$ and the phase of each component varies 
diagonally as shown in Fig.~\ref{soc_state}(a). Further increase in $J$ 
results into a transition to zFM superfluid. The effects of SOC at higher 
strengths $\lambda=0.03 U$ and $\lambda=0.04 U$ are shown in 
Figs.~\ref{pd_1p5}(c) and (d). At $\lambda=0.03 U$, the MI(3) completely 
vanishes, and the parameter regime of PT superfluid phase is enhanced. This is 
also evident from the SF region between the MI(1) and MI(2) lobes. At 
$\lambda=0.04 U$, the destruction of Mott lobes is enhanced as indicated by the 
absence of MI(2). As SOC strength increases, the melting of insulating lobes 
occurs first for higher density lobes and then it continues to the lower ones. 
In addition, the $J_c$ of MI-SF transition also decreases with $\lambda$. 
We find that at higher $\lambda\approx 0.065 U$, the MI(1) phase 
gets completely destroyed and the system exhibits superfluid phase transition 
between PT and zFM states. The transition between the PT and zFM is a broad 
one. To get the phase boundary for this transition, we have used the nonlinear 
least squares fitting and the residuals is of the order of $10^{-3}$. It is 
important to note that for stronger interspin interaction, 
$U_{\uparrow\downarrow}/U > 1$, we do not observe the ST phase. This is 
consistent with the quantum phases of continuum system with SOC where the 
tuning of Raman coupling for strong interspin interaction does not lead to ST 
phase~\cite{chen_17}. 


\subsection{Finite temperature results for $U_{\uparrow\downarrow} = 0.5$}
\begin{figure}[ht]
  \includegraphics[width=0.9\linewidth]{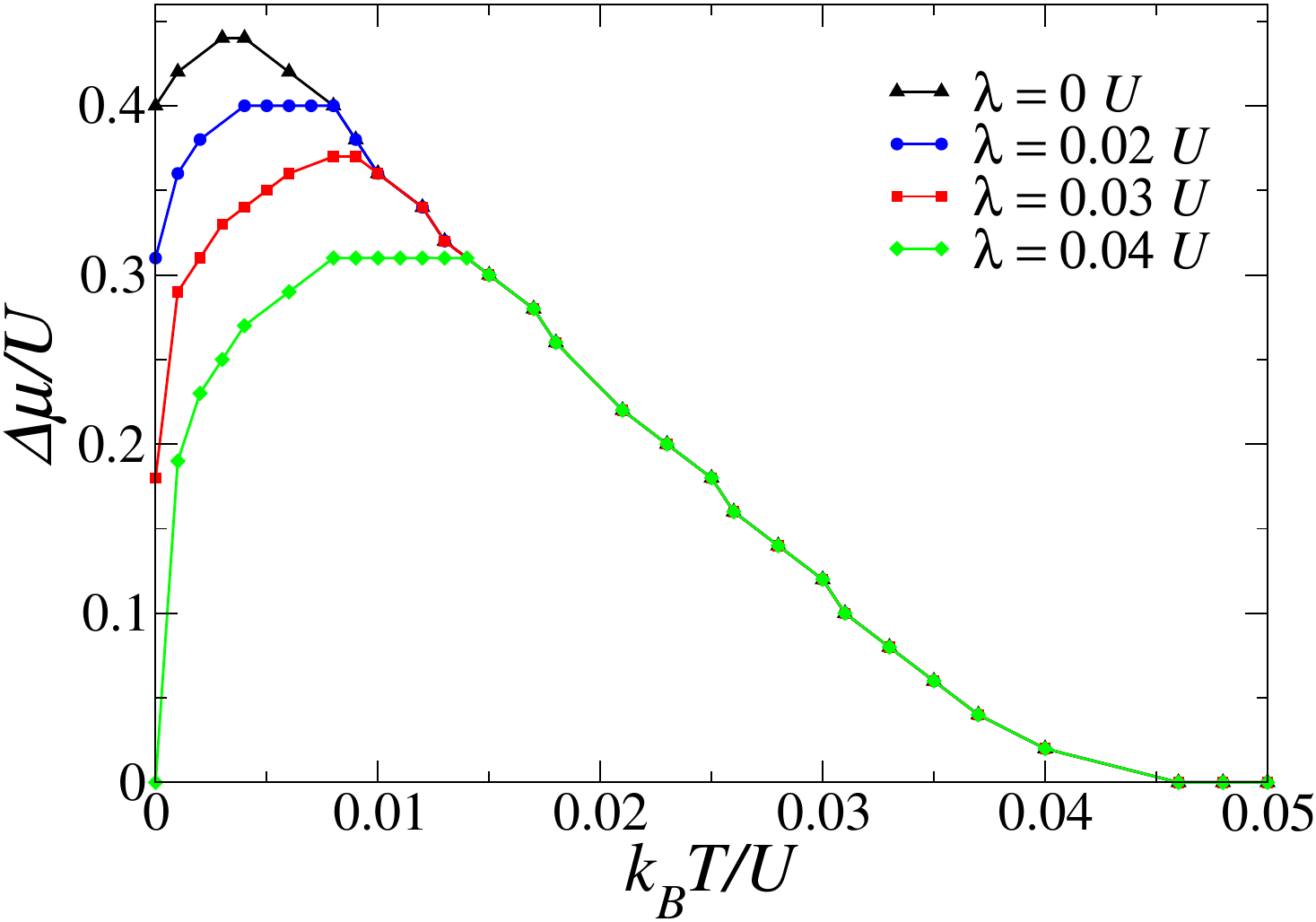}
  \caption{The width of first Mott lobe MI$(1)$ at $J = 0.01 U$ as a
           function of temperature for various SOC strengths. The values of
           $\lambda$ in units of $U$ are shown in the upper right corner of
           the plot. Here, the interspin interaction 
	   $U_{\uparrow\downarrow} = 0.5U$. At lower temperatures, the melting
           of the MI lobe depends on the value of $\lambda$, and at higher
           $k_{B}T$ the width remains similar to $\lambda=0$ case.
           For all cases, with and without SOC, the MI$(1)$ phase completely
           melts and replaced by NF phase at $k_{B}T/U\approx 0.046$.}
 \label{mi_soc}
\end{figure}

  At finite temperature, the Mott lobe melts into NF phase due to thermal 
fluctuations. The NF phase has no long-range order, but it is compressible 
($\kappa\neq0$). Therefore, this phase can be distinguished from the 
insulating MI phases by finite $\kappa$. We examine the melting of MI phase as 
a function of temperature for various SOC strengths. In Fig.~\ref{mi_soc}, we 
plot the width of the first Mott lobe MI(1) at $J = 0.01U$ for different 
values of $\lambda$. For $\lambda=0$, at lower temperatures, the width of the 
MI lobe first increases for $k_{B}T/U < 0.004$. 

At $k_{B}T/U\approx0.004$ the Mott lobe starts melting and width of the lobe 
decreases with temperature. At $k_{B}T/U\approx0.046$, the MI phase is 
completely replaced by NF phase. However, for finite $\lambda$, there is a 
combined effect of SOC and finite temperature on the width of the Mott lobe. 
For $\lambda = 0.02 U$, at low temperatures the width first increases then at 
$k_{B}T/U\approx0.009$ the thermal fluctuations overcome the SOC effects, and 
this leads to decrease in the width. The effects of SOC are prominent at 
larger values of $\lambda$ as evident for $\lambda = 0.03 U$ and $0.04 U$ 
cases in Fig.~\ref{mi_soc}. At higher temperatures, the melting of MI phase 
is independent of $\lambda$, and the decrease in the width of MI lobe is 
similar to $\lambda=0$ case. 
\begin{figure}[ht]
  \includegraphics[width=\linewidth]{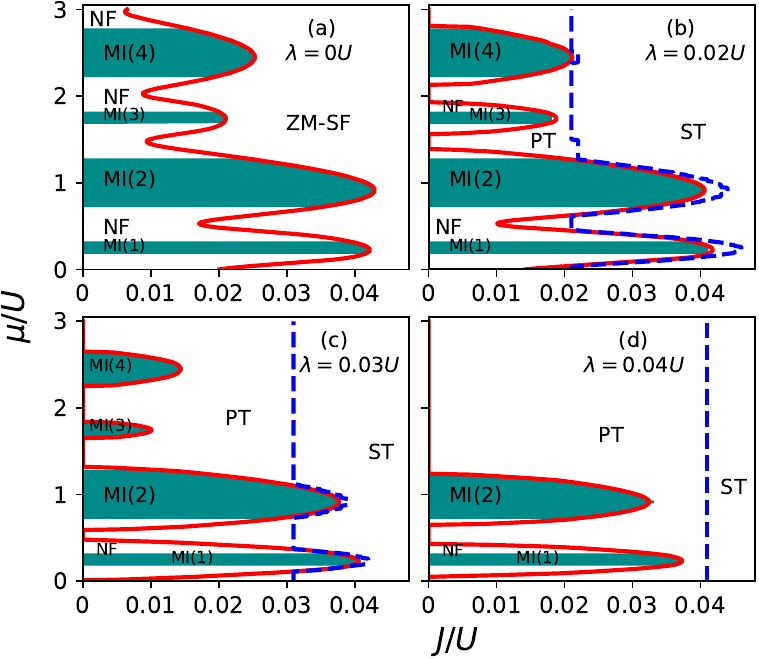}
  \caption{The finite temperature phase diagram of BHM for different values of
           $\lambda$ at $k_{B}T/U=0.03$. The interspin interaction
           $U_{\uparrow\downarrow} = 0.5U$. The shaded green bands are the 
           insulating MI regions which is distinguished from NF phase present 
	   at finite $T$. The reemergence of insulating regimes and 
	   destruction of PT superfluidity at finite temperatures are observed.
	   The constant width of MI$(1)$ for both zero and finite SOC confirms 
	   the behaviour reported in Fig.~\ref{mi_soc}. The blue dashed line 
           represents the PT-ST superfluid phase transition obtained using 
           finite-temperature Gutzwiller mean-field approach.}
 \label{bhm_soc_T}
\end{figure}

We further discuss the finite temperature phase diagram at $k_{B}T/U=0.03$, 
and it is shown in Fig.~\ref{bhm_soc_T}. At $\lambda=0$, the thermal 
fluctuations destroy the off-diagonal long-range order of SF phase, and extend 
the parameter space of NF phase. Odd Mott lobes stretch along $J/U$ axis. For 
example, at $k_{B}T=0$, the critical hopping of MI(1)-SF transition is 
$0.0268$ in Fig.~\ref{bhm_soc}(a) and at $k_{B}T/U=0.03$ it increases to 
$0.0422$ as evident from Fig.~\ref{bhm_soc_T}(a). Similar enhancement is also 
apparent for MI(3) from the comparison of Figs.~\ref{bhm_soc_T}(a) and
~\ref{bhm_soc}(a). In the presence of SOC, there is an interplay of the effects
of SOC and finite temperature. For smaller SOC strengths ($\lambda = 0.02$ and 
$0.03$), the remarkable feature of the reemergence of MI lobes at the cost of 
finite-momentum superfluids at finite temperature is observed. In particular, 
the SOC-induced PT phase near the atomic limit in Fig \ref{bhm_soc}(b) melts 
into MI(3) and NF phases as shown in Fig.~\ref{bhm_soc_T}(b). The destruction 
of PT superfluidity with a wide region of NF state at finite temperatures is 
consistent with the previous Monte Carlo study of strongly-correlated bosons 
with SOC~\cite{hickey_14}. The reemergence of insulating domains at finite 
temperature is in agreement with our analysis of the width of MI lobe with SOC 
which is shown in Fig.~\ref{mi_soc}. While increasing the SOC strengths from 
$0.02$ [Fig.~\ref{bhm_soc_T}(b)] to $0.03$ [Fig.~\ref{bhm_soc_T}(c)], the PT 
state is favoured by melting the NF and MI phases. At $\lambda=0.04 U$, the 
phase boundary of PT to ST superfluid transition, as in the case of zero 
temperature, remains unchanged and independent of the average particle 
densities or $\mu$ at finite temperatures.

\begin{figure}[ht]
  \includegraphics[width=\linewidth]{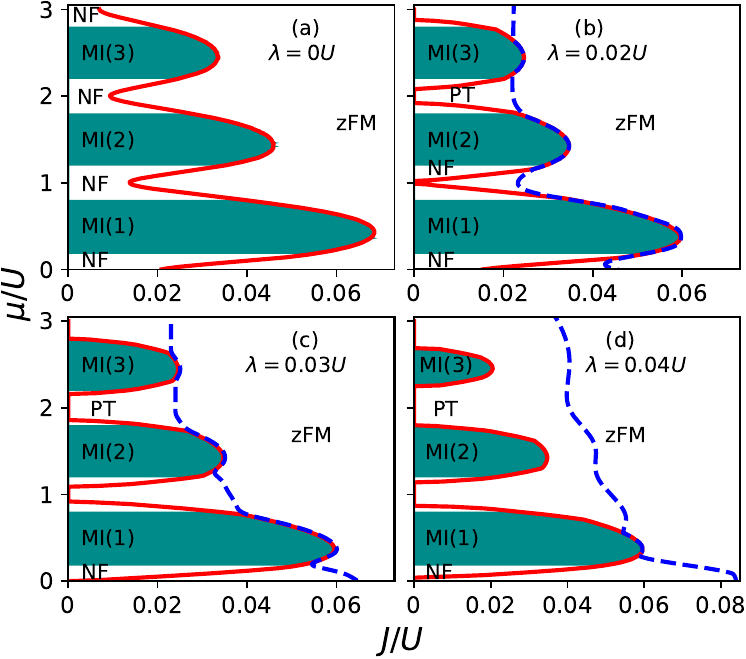}
  \caption{The finite temperature phase diagram of BHM for different values of
           SOC strengths at $U_{\uparrow\downarrow} = 1.5U$. The shaded green
	   bands represent the residual insulating domains in the presence of 
	   thermal fluctuations. Outside of the bands tiny white region shows 
	   NF phase. The thermal energy corresponding to temperature is 
	   $k_{B}T/U=0.03$. The finite temperature stabilizes the MI phases
	   against the SOC and suppress the finite-momentum superfluidity. 
	   This is evident from (c,d), as compared to corresponding zero 
	   temperature case shown in Figs.~\ref{pd_1p5}(c,d).}
\label{pd_1p5_KBT}	   
\end{figure}

\subsection{Finite temperature results for $U_{\uparrow\downarrow} = 1.5$}
Like in the case of $U_{\uparrow\downarrow}=0.5 U$, we examine the 
finite-temperature phase diagram in the phase-separated regime with SOC. In 
particular, we explore the stability of the finite-momentum superfluids with 
the thermal fluctuations arising from finite $T$. To gain better insights, we 
briefly review the $\lambda=0$ case. As expected the insulating lobes melt to 
the NF phase at $k_{B}T/U=0.03$, and this is discernible from the phase 
diagram shown in the Fig.~\ref{pd_1p5_KBT}(a). The phase diagrams of 
$\lambda\neq 0$ at $k_{B}T/U=0.03$ are shown in the 
Fig.~\ref{pd_1p5_KBT}(b)-(d). At $\lambda=0.02 U$, the thermal fluctuations 
favour the insulating domains, and the emergence of NF phase between the MI 
lobes reduces the PT superfluidity as is evident from the comparison of 
Fig.~\ref{pd_1p5_KBT}(b) with Fig.~\ref{pd_1p5}(b). The reemergence of MI 
lobes at finite $T$ is also there in this case as can be confirmed by comparing 
Figs.~\ref{pd_1p5_KBT}(c)-(d) with~\ref{pd_1p5}(c)-(d). At the phase 
boundaries, the critical hopping of the MI-zFM and PT-zFM transitions are 
shifted to higher $J$ at finite temperature as compared to the critical $J$ at 
zero temperature. The effect of increase in SOC strength leading to the 
increase in PT superfluid phase is also evident in the phase diagrams shown in 
the Figs.~\ref{pd_1p5_KBT}(c)-(d) similar to $U_{\uparrow\downarrow}/U < 1$ 
case. 


\section{Conclusions}
\label{conc}
 We have studied the parameter domain of various finite-momentum superfluids of
spin-orbit coupled ultracold bosonic atoms in two-dimensional optical lattices.
To examine various superfluid states with different atomic densities and phase 
ordering, we have used spin-dependent momentum distributions, a routinely 
measured observable in cold-atom experiments. For $U_{\uparrow\downarrow} < U$, 
with $\lambda/J \lessapprox 1$ the favored superfluid phase is PT, whereas with 
$\lambda/J \gtrapprox 1$ the system is in ST phase. Starting with PT phase, the 
increase in $J$, results in PT to ST superfluid phase transition. We have 
further shown that the inclusion of the quantum fluctuations via cluster 
Gutzwiller approach results in $\rm{ST_{den}}$ phase corresponding to the 
parameter domain of ST phase obtained with Gutzwiller mean-field theory. In the 
limit, $U_{\uparrow\downarrow} > U$, the stripe superfluid is absent, and 
phase-twisted to $z$-polarized ferromagnetic transition is observed as $J$ is 
varied. We have further shown that the thermal fluctuations destroy the 
phase-twisted superfluidity, and favour the insulating and normal states. 
The results of the present study are pertinent to the ongoing quantum gas 
experiments with spin-orbit coupling and offer a parameter space in the 
$J-\mu$ plane to observe the novel finite-momentum superfluids.


\begin{acknowledgments}
 We thank Deepak Gaur and Hrushikesh Sable for valuable discussions and 
acknowledge the support of High Performance Computing Cluster at IAMS, 
Academia Sinica, Taiwan. S.G. thanks Science \& Engineering Research Board 
(SERB), Department of Science and Technology, Government of India 
(Project: ECR/2017/001436) for support. 
\end{acknowledgments}

\appendix*
\setstretch{}
\section{Finite-temperature Gutzwiller mean-field theory}
 We incorporate the effects of the thermal fluctuations at finite temperatures
by considering the thermal average of the observable quantities. To compute 
the thermal average, we first get the full set of eigenspectrum obtained from 
the diagonalization of the mean-field Hamiltonian. We further use the 
single-site energy spectrum $E^l_{i}$ and the eigenstates $\ket{\psi}^l_{i}$ 
to evaluate the partition function of the system
\begin{equation}
  Z_{i} = \sum_{l=1}^{N_b}e^{-\beta E^l_{i}},
\end{equation}
where $l$ is the eigenstate index, $N_b$ is the Fock space dimension, 
$\beta = (k_{B}T)^{-1}$, and $T$ is the temperature of the system. At finite 
$T$, the region of the phase diagram with vanishing SF order parameter and the 
real number occupancy $\langle \hat{n}_{i\alpha} \rangle$ is defined as the 
normal-fluid state. 

From the definition of the partition function, the thermal average of the SF 
order parameter is
\begin{equation}
  \langle \phi_{i\alpha}\rangle = \frac{1}{Z}\sum_{l=0}^{N_b}
                            \prescript{l}{i}{\bra{\psi}}
			    \hat{b}_{i\alpha} e^{-\beta E^l} \ket{\psi}^l_{i},
\end{equation}
where $\alpha={\uparrow,\downarrow}$ is the spin-component index and 
$\langle\cdots\rangle$ represents the thermal averaging of $\phi$. 
Similarly, the atomic occupancy at finite $T$ is defined as
\begin{equation}
  \langle\langle \hat{n}_{i\alpha} \rangle\rangle = \frac{1}{Z}\sum_{l=0}^{N_b}
                          \prescript{l}{i}{\bra{\psi}}
			  \hat{n}_{i\alpha} e^{-\beta E^{l}}\ket{\psi}^l_{i}.
\end{equation}
The average occupancy is 
$\langle n_{\alpha} \rangle = \sum_{i}\langle\langle \hat{n}_{i\alpha} 
\rangle\rangle /L$. At finite $T$, the spin-dependent momentum 
distributions $\langle \rho_{\uparrow,\downarrow}(\mathbf k) \rangle$ are 
computed from the thermal-averaged SF order parameters.

\bibliography{so_opl}{}
\bibliographystyle{apsrev4-1}
\end{document}